# Estimating the quality of academic books from their descriptions with ChatGPT[1]

Mike Thelwall, Andrew Cox: Information School, University of Sheffield, UK.

Although indicators based on scholarly citations are widely used to support the evaluation of academic journals, alternatives are needed for scholarly book acquisitions. This article assesses the value of research quality scores from ChatGPT 4o-mini for 9,830 social sciences, arts, and humanities books from 2019 indexed in Scopus, based on their titles and descriptions but not their full texts. Although most books scored the same (3* on a 1* to 4* scale), the citation rates correlate positively but weakly with ChatGPT 4o-mini research quality scores in both the social sciences and the arts and humanities. Part of the reason for the differences was the inclusion of textbooks, short books, and edited collections, all of which tended to be less cited and lower scoring. Some topics also tend to attract many/few citations and/or high/low ChatGPT scores. Descriptions explicitly mentioning theory and/or some methods also associated with higher scores and more citations. Overall, the results provide some evidence that both ChatGPT scores and citation counts are weak indicators of the research quality of books. Whilst not strong enough to support individual book quality judgements, they may help academic librarians seeking to evaluate new book collections, series, or publishers for potential acquisition.

## Introduction

Academic librarians sometimes need to use quantitative indicators to help them judge the quality of academic research, although they are not recommended as the sole basis for decision making (Hicks et al., 2020; Wilsdon et al., 2015). For example, academic librarians might check Journal Impact Factors or similar citation-based indicators or rankings when making journal subscription decisions or recommendations (Hoffmann & Doucette, 2012). Book-based indicators might be useful for academic librarians in charge of bulk acquisition decisions or recommendations, excluding patron-driven acquisitions (Tyler & Boudreau, 2024). Whilst book usage has been extensively studied (e.g., Walker, 2021), estimating the value of books in advance of purchase is important for subscription decisions. Supporting information could be particularly useful for large e-book collection subscriptions (Zhang, 2020). Because of this, there has been some interest in citation data for some types of book (Hoffmann & Doucette, 2012).

The use of citation data for journals is possible for Western nations because their primary journals tend to be systematically indexed by databases like Dimensions, Scopus, and the Web of Science. It is also possible for scholars in areas that have formed equally effective local databases, such as SciELO (Vélez-Cuartas et al., 2016). The situation is different for books, however. Although they are longer than articles and therefore more time consuming to thoroughly evaluate, especially for entire collections, citation-based book indicators are not systematically available. Whilst Scopus and the Web of Science index books from selected academic publishers, their coverage seems to be far from complete (e.g., Torres-Salinas et al., 2014). This causes two problems. First, any evaluation is likely to find that most books have no record and therefore no citation counts. Second, since citations to books often originate from other books, the citation counts for books in Scopus and the Web of Science are likely to be incomplete and biased against book-oriented fields.

There have been several attempts to generate useful citation data for books, primarily driven by their importance in the arts and humanities and some social sciences. Evaluations of book citations in the Web of Science found that edited collections were more cited than monographs, perhaps by counting citations to the individual chapters as well as the whole book. Books from academic presses also tended to be more cited. This dataset included many edited volumes as well as books in series (Torres-Salinas et al., 2014). Alternatively, carefully constructed queries can be used to identify citations to books from Google Books (Kousha & Thelwall, 2015). This finds 50% more citations to books than Scopus, but fewer than Google Scholar (Kousha et al., 2011). Searching Google Books only works for books with reasonably distinctive titles, however, and it seems likely that there will be a lag between books being published and their full text being indexed by Google Books, if the publisher allows it. Google Scholar is another option but is difficult to use for large sets of books because searches with it can only be automated to a limited extent. The OpenCitations project is also a major source of book citations (Zhu et al., 2020).

There is a little empirical information about citation rates for books. An analysis of citations to social science books, counted with Google Books, suggested that they tend to attract three times as many citations as articles, but that US politics is a relatively uncited area (Samuels, 2013). There is also some evidence about the relationship between book citation counts and other indicators. For example, Scopus citation counts to textbooks correlate weakly and positively with their uptake in academic syllabi (Maleki et al., 2024). In contrast, books in the humanities and social sciences have very weak and often negative correlations with common altmetrics (Yang et al., 2021). Despite these findings, no prior study seems to have investigated the relationship between citation counts and book quality.

A new alternative to citation data is the use of Large Language Models (LLMs) to score documents for quality based on their text or short summaries. For journal articles, ChatGPT 4o-mini can score articles for research quality on a four-point scale, with results that correlate positively with an indicator of expert quality judgements for all areas of science (Thelwall & Yaghi, 2024; Thelwall et al., 2024). For this, the full text is not needed and, in the limited experiments so far, the optimal input is the article title and abstract without the remaining text (Thelwall, 2024ab; Zhou et al., 2024). Whilst a criticism of citation-base indicators has been that they primarily reflect scholarly impact rather than other important dimensions of research quality, including societal impact, rigour, and significance (Langfeldt et al., 2020), LLMs can be explicitly asked to consider all quality dimensions. Of course, since the optimal results do not require full texts, the LLM scores are guesses rather than evaluations at any level. They primarily seem to leverage author claims in abstracts with some world context from the LLM training corpus.

Given the limitations of citation data for books and the promise of LLMs for academic journal evaluation, the goal of this study is to investigate whether ChatGPT can meaningfully assess the quality of academic books, particularly to support book collection development. Since there is no gold standard for book quality and no list of books annotated for research quality, the sample analysed will be books with citation data so that the two types of scores can be compared in a validation test for ChatGPT. The following research questions drive the work. The focus is on the social sciences and arts and humanities, where books are more important than in the physical and health sciences and engineering.

- **RQ1**: Do social sciences and arts and humanities books with higher ChatGPT quality scores also tend to be more cited? A positive answer would indicate that neither is random and suggest that one is an indicator of research quality (however weak) if the other is.
- **RQ2**: Which types of books does ChatGPT give high or low scores to? And is this different to the types of books that attract many or few citations?

## Methods

The research design was to obtain a large sample of scholarly books from the arts, humanities, and social sciences, use ChatGPT to estimate their quality and compare the results with citation counts. The secondary method, for RQ2, was to identify themes in books that attract high or low ChatGPT scores to get insights into the factors influencing scores and hence their validity.

### Dataset

Although there are relatively comprehensive book sources related to ISBNs, for this article the relevant texts are purely academic works. The most appropriate source seems to be the Scopus database, which apparently indexes more books than other citation indexes. For example the Web of Science book citation index includes 151,000 (Clarivate, 2025). Scopus is a suitable source because all books seem to be scholarly, and they are classified by broad subject area.

Social science books in Scopus were obtained on 3 November 2024 by submitting the query: SUBJMAIN(3300) AND DOCTYPE(bk) with the year specified as 2019, where 3300 is the All Science Journal Classifications code for Social Sciences (all). The year 2019 gives a nearly five-year citation window to ensure that the citation data is reasonably mature, even accounting for the slower rate of citation for books (Wang, 2013). This was repeated for all social sciences categories from 3301 to 3322 and for the arts and humanities categories 1200 to 1213. Almost all the books were in the general categories 3300 Social Sciences (all) and 1200 Art and Humanities (all), however, so the few books in the other categories (e.g., one book in 3304 Education) were merged into one social sciences and one arts and humanities category.

For ChatGPT a book description was needed to base its evaluation on. In Scopus, some books did not have descriptions and in other cases they were short. After examination of the data, a minimum of 750 characters was selected as a threshold for an adequate size description. All books with shorter descriptions were removed.

Books in Scopus can have multiple subject categories and many of the books found in both arts and humanities and social sciences categories. These were split into a separate category, so the final dataset was in three non-overlapping parts: social sciences only, arts and humanities only, and dual category books.

### ChatGPT scores

The Scopus data includes an "Abstract" field containing an author or publisher (e.g., marketing, managers, editors) description of the book contents. Depending on the publisher, it might derive from an author abstract within the book, or some or all of the back cover blurb (Bacic, 2021). A check of 10 randomly selected books from the dataset found that these descriptions were always either back cover blurbs or (for e-books, which typically lack a back cover) in a paragraph performing the role of an abstract and written like a blurb in the book half title page. The Scopus abstracts usually excluded the endorsement quotes from other scholars that are usually found underneath blurbs on back covers, however. Academic blurbs typically describe the book, demonstrate the author's expertise, highlight key contents, and evaluate the book implicitly by stating target audiences or explicitly through endorsement quotes (Cacchiani, 2007).

Previous research has suggested that for optimal results from ChatGPT for quality scores of journal articles, only the titles and abstracts should be submitted and not full texts (Thelwall, 2025). It was therefore reasonable to try the same for books. Ideally, this should be repeated for book full texts to check if the advantage of abstracts for articles replicates for books, but only book descriptions were available. In the UK, where the research took place, it is legal to use lawfully accessed texts for data mining (Headon, 2023) and the ChatGPT API used does not retain the submitted data for training (openai.com/enterprise-privacy), so there is no potential copyright violation. In other countries, the legality of this approach should be checked before use.

Each book title/abstract was submitted to ChatGPT 4o-mini following the prompt, "Score the following book:". No other information about a book was entered. For example, its author, publisher and citation rate were not included. The prompt was accompanied by system instructions describing the Research Excellence Framework (REF) [2] criteria for academic research taken from the level definitions or the four "Main Panels" in the official REF guidelines (REF, 2019). The REF is the UK's periodic systematic process to evaluate the research of government funded institutions, with the results used to determine the amount of research funding given to each one in the next period. It includes a year of reviewing research outputs and other documentation by panels of over 1000 experts (REF, 2023). The REF guidelines for social science research (Main Panel C) and arts and humanities research (Main Panel D) were used (the exact system instructions are here: Thelwall & Yaghi, 2024). These were matched with the appropriate categories for two of the datasets and the social sciences guidelines were used for the combined dataset. From comparisons of the two guidelines, the social sciences guidelines seemed more appropriate to the combined set, but this was a subjective decision. The ChatGPT API was used for the queries during 4-5 November 2024 and the entire set was submitted five times. Averaging the results of multiple iterations increases accuracy (Thelwall, 2024, 2025), and after five iterations the advantage of each extra iteration is small.

The result of the above is a set of ChatGPT quality reports on individual books, each accompanied by a score using the REF guidelines: 1* (nationally relevant research), 2*, 3*, or 4* (world leading). The REF guidelines used comprise explanations of three quality components: rigour, originality, and significance. The reports therefore sometimes scored these separately without an overall score. In such cases, the overall score would be taken to be the average of the three individual scores (to three decimal places). The reports also sometimes included fractional scores (e.g., 2.5*) and these were accepted. The scores were automatically extracted from the reports by a program written for this task in Webometric

---

Analyst[3]. This program used the score that it had extracted except that it asked the first author of this article in cases when no extraction rule had been triggered.

## Citation data

Log-transformed citation counts ln(1+citations) were used as the citation data (Thelwall, 2017), without field or year normalisation. The log transformation removes most the extreme skewing found in citation datasets, with 1 added because the log of zero is undefined. In cases where a logarithm needs to be applied to a collection of numbers including zeros, it is standard practice to add a small amount to each number to make all numbers positive so that they can then all be logged. The number 1 is the default value to be added, including for citation analysis. Field and year normalisation is usually employed with citation data, but was not necessary case because almost all books in each dataset were from a single (or the same two for the dual set) Scopus narrow field(s) and year.

## WATA

Word Association Thematic Analysis (WATA) (Thelwall, 2021) was used to identify themes in the high and low scoring books. Although other text analysis methods could have been used instead of WATA, its advantage is its statistical basis to guard against spurious patterns being reported. It works by decomposing each book title and description into its constituent words, then finding individual words that occur disproportionately often in the half of the books with a higher ChatGPT score compared to books with a below average ChatGPT score. A 2x2 chi squared test was used for each word in the list, supported by a Benjamini-Hochberg familywise error correction procedure (Benjamini & Hochberg, 1995) to reduce the chance of spurious positive results from conducting many simultaneous tests. The result of this was a list of words occurring disproportionately often in high scoring book titles and descriptions.

 The statistically significant words were then examined to find their typical meaning and use context in the dataset by reading ten randomly-selected titles and descriptions around the given term (i.e., the key word in context approach) and by examining the terms that most frequently co-occur with it. For example, the term "framework" was allocated to the theory theme since it usually occurred in the phrase "theoretical framework" and the term "argue" was added to the analysis theme since it was used to introduce an analysis offered by the book.

 After the statistically significant words were contextualised as above, they were iteratively clustered into themes of related words. The easiest cases were when two words were in a common phrase, but in other cases a judgement had to be made about whether the contexts of the words were similar enough to class them as a common theme. This process was iterative and involved additional context checking to clarify uncertainties. The result of this was a set of themes associated with high scoring books.

 The above was repeated for low scoring books, for books cited above average, and for books cited below average, giving four sets of themes. The high cited and high scoring themes were developed in parallel to ensure as much overlap as possible, and the same for the low cited and low scoring themes. The entire process was conducted separately for the social sciences dataset, the arts and humanities dataset, and the dual dataset giving (in theory, but see below) twelve sets of themes. Finally, themes were compared between datasets to check whether they could align to simplify the results.

---

[3] github.com/MikeThelwall/Webometric_Analyst

# Results

## *Comparison with citations*

The median ChatGPT score for all three datasets is 3*, with all except five scores being between 2* and 4*. For all three datasets, there is a statistically significant moderate positive Pearson correlation between log-transformed citation counts and ChatGPT scores (Table 1). Thus, assuming that one of the two is a quality indicator, this gives evidence that the other also is. Whilst there is not strong evidence in either case, the correlations show that the two are related and provide supportive evidence for both being valid, albeit possibly weak, research quality indicators.

Table 1. Pearson correlation between ChatGPT 4o-mini scores (average of five scores per book) and log-transformed citation counts for Scopus-indexed books published in 2019.

| Logged citations vs. ChatGPT score | Social sciences only | Arts and humanities only | Arts and humanities and social sciences |
|---|---|---|---|
| Pearson correlation | 0.377 | 0.358 | 0.383 |
| 95% confidence interval | [0.351, 0.403] | [0.331, 0.385] | [0.341, 0.423] |
| Sample size | 4112 | 4041 | 1677 |

For all three datasets, 3* is by far the most common score. For social sciences books, the positive correlation between (logged) citation counts and average ChatGPT scores occurs because books with citation counts above 100 never score below 3*, whereas books with higher citation counts have a small chance of scoring above 3* (Figure 1). Thus, the positive statistical association between citation counts and ChatGPT scores seems to be primarily due to the occasionally higher ChatGPT scores for highly cited books. There is a similar pattern for arts and humanities books with a cut-off of 150 citations (Figure 2) and for dual category books with a citation count threshold of 200 (Figure 3).

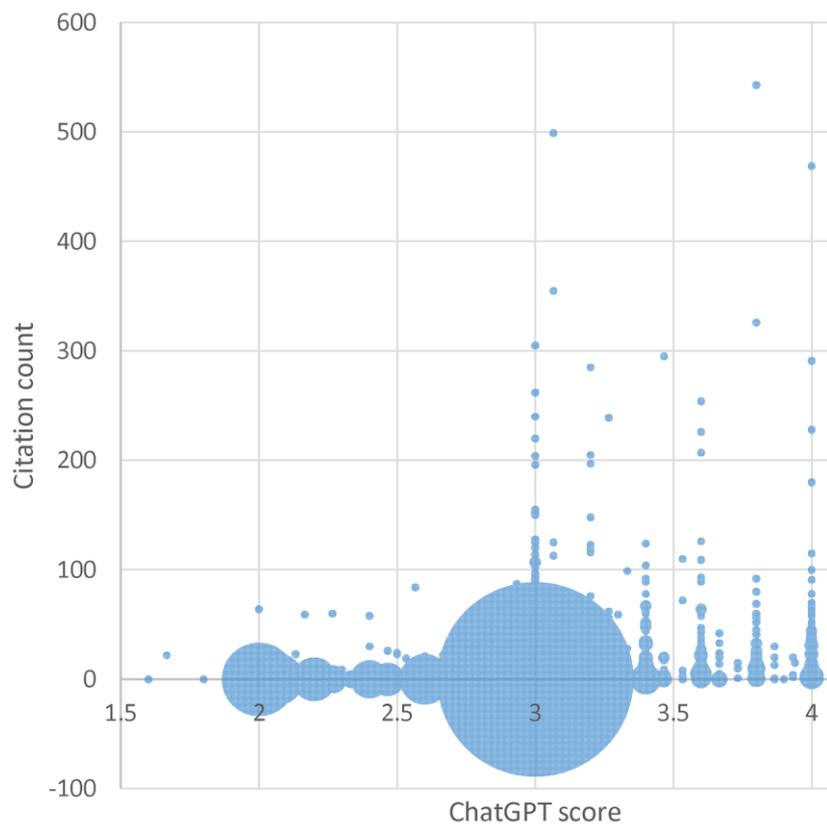

Figure 1. Citation counts against average ChatGPT 4o-mini score for 4112 social sciences books published in 2019.

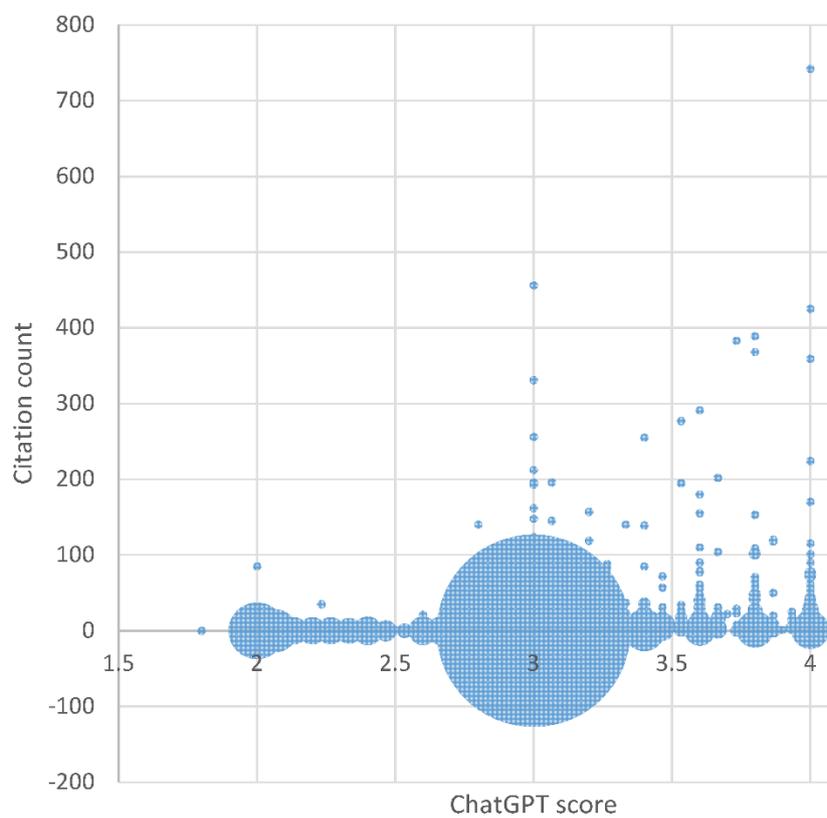

Figure 2. Citation counts against average ChatGPT 4o-mini score for 4041 arts and humanities books published in 2019.

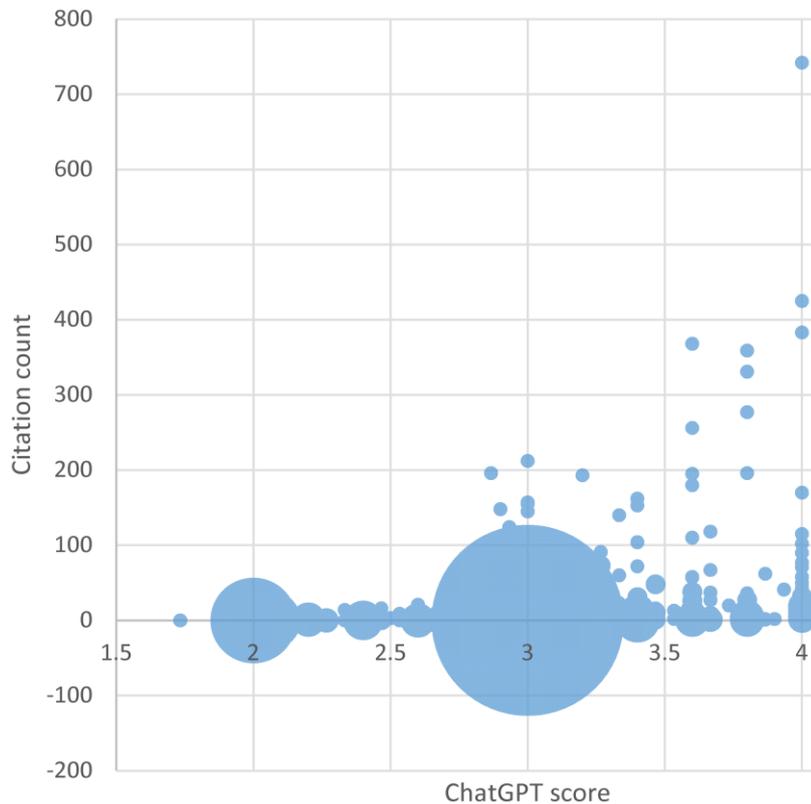

Figure 3. Citation counts against average ChatGPT 4o-mini score for 1677 arts and humanities and social sciences books published in 2019.

## *Articles with high ChatGPT scores and/or citation rates*

Many social sciences book themes were identified with WATA (Table 2) and these are summarised below in five broad groups.

**Topics**: Some themes are topic-specific in the sense that books about the topic tend to be more cited and/or get higher ChatGPT scores. Three topics that associate with high values on both: legal issues, politics, and governance. In addition, sociology research tends to get high ChatGPT scores.

**Audiences**: Some themes are associated with the stated intended audience of the book, with books for scholarly audiences scoring higher with ChatGPT and citations. In contrast, a student or practitioner audience associates with more citations but not with higher ChatGPT scores.

**Methods**: Whilst interviews and ethnography associate with higher ChatGPT scores, data and evidence associate with more citations. Whilst data suggests quantitative research, evidence could be of any type.

**Approach**: Book descriptions that explicitly mention theory, analysis and/or the wider context of the topic tend to be more cited and attract higher ChatGPT scores.

**Novelty**: Book descriptions that claim novelty attract more citations but not particularly higher ChatGPT scores. This is perhaps surprising since novelty is part of the instructions for ChatGPT but is not directly necessary for citations.

Table 2. Themes identified through WATA from terms in highly cited or high scoring books compared to low cited or low scoring books (social sciences).

| Theme | High ChatGPT terms | High citation terms |
|---|---|---|
| Terms reporting an analysis of the topic of the book | Explore, **how, drawing**, examine, argue, analysis, role, analyse, dynamic, demonstrate, between | **How, drawing**, offer, offering, understanding, goes |
| Theory used in the article | **Empirical**, theoretical, framework, **theory** | **Theory**, **empirical** |
| Setting a wider context for the research reported | **Across**, context, global | **Across** |
| A scholarly intended readership for the book | **Scholars, researchers, field** | **Scholars, researchers, field** |
| Legal issues | Legal, **justice**, right | **Justice** |
| Politics, economics and democracy | Political, politics, (international) relations | (non-state) actor |
| Governance, policy, and accountability | **Governance** | **Governance** |
| The current relevance of the research is mentioned. | Contemporary | |
| Sociology | Sociology | |
| Ethnographic study | Ethnographic | |
| Interviews reported in book | Interviews | |
| Data analysed in the book | | Data |
| Education and professional practice | | Practice |
| Evidence presented in the book | | Evidence |
| The book targets students and practitioners | | Student, practitioner |
| Claim for the book's originality | | Innovative, novel |

**Topics**: For humanities books (Table 3), high scoring topic-specific themes include cultural or social history (ChatGPT) and language/linguistics (citations).

**Audiences**: Humanities books for scholarly audiences tend to score higher with ChatGPT without necessarily attracting more citations.

**Approach**: Books that explicitly mention analysis tend to be more cited and attract higher ChatGPT scores, and those that mention a wider context tend to be more cited.

Table 3. Themes identified through WATA from terms in highly cited or high scoring books compared to low cited or low scoring books (arts and humanities).

| Theme | High ChatGPT terms | High citation terms |
|---|---|---|
| Terms reporting an analysis of the topic of the book | Explore, **how, argue**, analysis | **How, argue** |
| Cultural or social history | Cultural | |
| A scholarly intended readership for the book | Scholars | |
| Language and linguistics | | Linguistics, applied |
| Setting a wider context for the research reported | | Across |

**Topics**: For books that are both humanities and social sciences (Table 4), topic-specific themes include a named area of studies (ChatGPT) or language/linguistics (citations).

Table 4. Themes identified through WATA from terms in highly cited or high scoring books compared to low cited or low scoring books (social sciences and arts and humanities).

| Theme | High ChatGPT terms | High citation terms |
|---|---|---|
| Topic area (x studies) | Studies | |
| Language and linguistics | | Linguistics, language |

## *Articles with low ChatGPT scores and/or citation rates*

A different range of themes was identified for books attracting few citations or low ChatGPT scores (Table 5).

**Topics**: For social sciences books, historical topics tended to be less cited and receive lower scores. Therapy research is scored lower by ChatGPT and interdisciplinary, military, foreign policy, and US-centred books tend to be less cited.

**Book type**: Textbooks and edited volumes tended to receive fewer citations and to attract lower ChatGPT scores.

**Audiences**: Books targeting wide audiences tend to receive lower ChatGPT scores.

**Style**: Descriptions in which the author thanks friends and colleagues tend to be scored lower by ChatGPT whereas books describing the author indirectly are less cited. In the former case, most of the 18 Scopus-indexed descriptions mentioning thanks were exclusively author acknowledgment texts and did not mention the book contents, giving ChatGPT little evidence for a research quality score (e.g., "Writing this book has been a rewarding but challenging experience. Along the way I have had the helpful comments and criticisms of a number of friends and colleagues. [] read parts of it. They and the reviewers for [] have weeded out many errors and forced me to think through a number of difficult points. The errors and confusions that remain do so in spite of their best efforts. I thank them all. I would also like to thank [] for his support of this project and for his editorial suggestions, which resulted in major improvements. Finally, thanks to [] for her editorial assistance in preparing the manuscript for publication."). Thus, these descriptions seem to have been entered into Scopus by mistake since they do not describe the book.

Table 5. Themes identified through WATA from terms in low cited or low scoring books compared to high cited or high scoring books (social sciences).

| Theme | Low ChatGPT terms | Low citation terms |
|---|---|---|
| Book is a collection of papers | **Paper**, presented, **originally** | **Paper,** essay, volume, **originally** |
| Historical topic | **Had, year, was, soviet, plan** | **Was, year**, attempt, **had**, were |
| Book is a textbook | Your, you, skill, help, teacher, activities, resource, course, exam, SENCO, **session**, guidance, tip, college, topic | **Session** |
| Copyright statement | **Imprint** | **Imprint** |
| Book will help the reader stay up to date | **Stay** | |
| Author perspective about book, including thanks | **My, me, friend, grateful, gratitude, am** | |
| Broad intended audience | Any, professional | |
| Therapy | Counsellor | |
| Indirect perspective about author or bio | | Dr, he, analyse |
| Inter-disciplinary perspective | | Inter-disciplinary |
| Foreign policy | | Foreign, Reagan |
| Military-legal | | Defense, force |
| USA | | United, American |

A wide range of themes were found by the WATA for low cited or low scoring humanities books (Table 6).

**Topics**: For humanities books, librarianship and the USSR tended to be less cited and receive lower ChatGPT scores. Recreation, school-oriented research, and studies focusing on the work of others are scored lower by ChatGPT.

**Book type**: Textbooks, reprints, and edited volumes tend to attract lower ChatGPT scores. Books offering extra content (e.g., a glossary) also tend to get lower ChatGPT scores, as do small books and those that offer practical advice.

**Audiences**: Books listing potential readers tend to get lower ChatGPT scores.

**Style**: Books describing the author in the third person are less cited.

Table 6. Themes identified through WATA from terms in low cited or low scoring books compared to high cited or high scoring books (arts and humanities).

| Theme | Low ChatGPT terms | Low citation terms |
|---|---|---|
| Book related to librarianship | **Librarian**, libraries, library, information, service, electronic, budget | **Librarian** |
| The USSR | **USSR** | **USSR** |
| Textbook | You, your, learn, informative, cover, help, instructor, student, prepare, reader, guide, practical, instruction | |
| Book is a reprint | Originally (published) | |
| Small book | Concise | |
| Book includes extra contents | Bibliography, glossary, timeline, additional, include, chronology, user-friendly, sidebar | |
| Book contains practical advice | Advice, tip | |
| List of potential readers | Academician | |
| Topic related to recreation | Recreational | |
| Topic related to schools | School | |
| Book is a collection of papers | | Essays, volume, collection, articles |
| Book is about the work of others | | Work, his |
| Third person style | | Author |

A few themes were found by the WATA for low cited or low scoring books that are dual categorised as both arts and humanities and social sciences (Table 7).

**Topics**: For dual subject books books, librarianship tended to be less cited and receive lower ChatGPT scores.

**Book type**: Textbooks, short books, and those offering extra content tend to get lower ChatGPT scores, whereas edited volumes tend to be less cited.

**Audiences**: Books with descriptions expressing uncertainty about reader knowledge tend to get lower ChatGPT scores.

Table 7. Themes identified through WATA from terms in low cited or low scoring books compared to high cited or high scoring books (social sciences and arts and humanities).

| Theme | Low ChatGPT terms | Low citation terms |
|---|---|---|
| Libraires and information services | **Librarian, library**, libraries, **published, service**, (library) manager | **Librarian, library, published, service** |
| Textbook | Your, ready, tip, reader, instructor, guide, resource, course | |
| Extra contents in book | Glossary, bibliography | |
| Book format | Format | |
| Small book | Concise | |
| Uncertainty about reader knowledge | May | |
| Book is a collection of papers | | Essay, collection, volume |

## Discussion

This article is limited by the sample of books analysed, which is not a complete or random sample of academic books and may be the result of commercial negotiations between Elsevier and various academic publishers for indexing rights. The descriptions contain some errors, such as acknowledgments instead of descriptions (18 cases), which may influence the results. The data also covers a single year and LLM, with newer and larger LLMs being likely to give better results. Moreover, if book description styles evolve then this might affect the patterns found here. A major limitation is that books were not classified into textbooks, monographs, and edited volumes, which would have allowed a finer grained analysis. Because of this, the positive correlation found could be entirely due to differing values and citation rates for the three different types of books. The nature of any ChatGPT biases is also unknown.

### RQ1: Comparison with previous research

The positive correlation between citations and ChatGPT scores based on titles and abstracts is not comparable to any prior book research although it echoes similar findings for social sciences, arts, and humanities journal articles (Thelwall & Yaghi, 2024; Thelwall et al., 2024). Although undermined by the inclusion of multiple book types, and particularly textbooks, the correlations above are much stronger and are more positive than correlations previously found with altmetrics (Maleki et al., 2024). Thus, they appear to be the statistically strongest evidence yet of the value of book citations as research quality indicators, whilst simultaneously tending to validate ChatGPT scores for the same role. These two intertwined points assume that there is a third factor that influences both but that is unrelated to research quality. They also assume, without evidence, that the correlation is not due to differing book types being analysed.

### RQ2: Analysis and comparison with previous research

In contrast to the results above, edited volumes have previously found to be more cited than monographs (Torres-Salinas et al., 2014), presumably because their citation counts often included the citations to individual chapters. Nevertheless, the results partly conflict with the relatively low citation rates for books about US politics (Samuels, 2013).

The WATA themes point to some plausible issues, although they do not seem to have been investigated explicitly before. First, the lower quality scores and fewer citations given to textbooks and teaching-oriented books may stem from their role being translational rather than introducing new research findings or analyses. Second, the lower citation rates for edited volumes may be a side effect of authors citing individual chapters rather than entire collections. There are no REF criteria for evaluating editorial contributions, however. Another uncontroversial issue is that concise books tend to be get lower ChatGPT scores, presumably because they are shorter or have a more summarising role rather than introducing new findings. Examples include, "Theory of spatial statistics: A concise introduction", "Interdisciplinary qualitative research in global development: A concise guide", and "A concise history of revolution". There are exceptions, such as "A concise history of Poland", which is moderately long (488 pages) with theoretical contributions (e.g., "examines the current issues facing a Poland which some would accuse of being out of touch with 'European values'."), but these seem to be the minority.

The lower ChatGPT scores and/or citation counts for some topics is consistent with previous research showing that citation rates (van Raan, 2004) and average ChatGPT scores (Thelwall & Yaghi, 2024) tend to vary between fields. Scopus may also index books with average quality levels that differ between fields. This might occur, for example, if it included a publisher that was the most prestigious for a topic. Similarly, the repeated occurrence of librarianship books in the lower categories might be due to Scopus indexing translational profession-oriented librarianship books rather than research monographs.

Books with descriptions mentioning theory, some methods and data tended to be more cited and/or get higher ChatGPT scores. Whilst the discussion of theory or analysis suggests a deeper contribution, evidence, data, or specific research methods suggest more substantial contributions. It seems reasonable that these would associate with more citations or higher scores. Alternatively, it may be a second order effect of these types of contents tending not to occur in textbooks or summarising books, or of disciplinary differences in average scores, with more analytical fields scoring lower and/or being cited less.

Some lower ChatGPT scores associate with styles used in their Scopus descriptions. Indirect styles (e.g., "the author shows") might detract from descriptions of the content and therefore give less scope for ChatGPT to detect research excellence. Alternatively, the style might be partly due to people other than the author writing the description.

## Conclusions

Although the results are not conclusive due to the absence of a set of recognised independent quality scores for books to benchmark citations and ChatGPT scores against, and the inclusion of textbooks and edited works within the collection, the results suggest that ChatGPT scores and citation counts have some power as research quality indicators in the social sciences and the arts and humanities. Because of these limitations, it is not clear whether they have practical value to help evaluate more uniform collections of books. In any case, the strength of both indicators seems weak and certainly too weak to consider when evaluating individual books. The main application might be for academic librarians to use ChatGPT to help judge the value of book series or collections that are being considered for acquisition. If the ChatGPT results are relatively clear cut, then this might help with decision making. The legal status with obtaining ChatGPT scores in terms of copyright should be checked first, however (Lucchi, 2024). In particular, in some countries it may not be legal to upload book titles and blurbs to machine learning systems. It may also breach copyright to upload blurbs to programs like ChatGPT that learn from their inputs and may recycle them for future outputs without any copyright acknowledgements.

The ChatGPT approach seems, in theory, to be more promising than citation counting because it can consider all quality dimensions. Nevertheless, it is currently limited by the need for book descriptions and the fact that the purpose of these descriptions often seems to be marketing, perhaps aligning with publisher styles. It would be useful to assess if ChatGPT gave better performance with book full texts. This might be possible in evaluative contexts where authors must submit electronic versions of their books for evaluation and permission can be obtained to enter them into a suitable LLM. It would be worth trying Google Gemini for this too, given its emphasis on multimodal information processing and the likely presence of images in many scholarly books.